\documentclass[]{aastex631}
\usepackage{blindtext}
\usepackage{amsmath}
\usepackage{comment}
\usepackage{bm}
\usepackage{soul}

\shorttitle{}
\shortauthors{}
\begin{document}

\title{Origin of the asymmetric gas distribution near the
co-orbital Lagrange points of an embedded planet}

\author[0009-0004-1149-9887]{Agustin Heron} 
\affiliation{Department of Astronomy, Indiana University, Bloomington, IN 47405, USA.}
\affiliation{Instituto de Astrofísica, Pontificia Universidad Católica de Chile, Av. Vicuña Mackenna 4860, 782-0436 Macul, Santiago, Chile.}
\author[0000-0003-0412-9314]{Cristobal Petrovich}
\affiliation{Department of Astronomy, Indiana University, Bloomington, IN 47405, USA.}

\author[0000-0002-3728-3329]{Pablo Benítez-Llambay} 
\affiliation{Facultad de Ingeniería y Ciencias, Universidad Adolfo Ibáñez, Av. Diagonal las Torres 2640, Peñalolén, Chile.}

\author[0000-0002-7056-3226]{Juan Garrido-Deutelmoser}
\affiliation{Department of Astronomy and Steward Observatory, University of Arizona, Tucson, AZ 85721, USA.}

\begin{abstract}
Hydrodynamic simulations of planet–disk interactions often show material accumulation near the co-orbital Lagrange points $L_4$ and $L_5$---features that may correspond to observed crescents in protoplanetary disks. Intriguingly, these simulations also show an asymmetrical distribution of gas between $L_4$ and $L_5$, whose physical origin is not yet understood and could allow to further constrain the inner workings of planet–disk interactions. We performed 2D hydrodynamic simulations of a single, non-migrating planet embedded in a gaseous disk to investigate this effect. We find that the asymmetry is solely controlled by the sign of the radial temperature gradient with positive gradients enhancing the accumulation at $L_4$ and negative ones enhancing $L_5$. A symmetric distribution is recovered on globally isothermal disks. Furthermore, we find that the azimuthal locations of $L_4$ and $L_5$ deviate from the classical circular restricted three-body problem, following a monotonic trend with the disk pressure scale height $h_{\rm p}$: $\phi_{\text{max}}\approx\pm60^{\circ}[1+0.18(h_{\rm p}^3M_\star/m_{\rm p})^{2/3}]$. Our simulations show that the longest-lived and largest-amplitude structures are produced by planets opening gaps with depths $\Sigma_{\rm gap}/\Sigma_0\lesssim 0.2$. We successfully reproduce the observed asymmetry using a semi-analytical model that incorporates the azimuthally asymmetric radial velocity background induced by the planet. Overall, our results suggest that asymmetries in the form of crescents and clumps inside of density gaps opened by planets can constrain the local thermodynamic properties of protoplanetary disks.
\end{abstract}

\section{Introduction} \label{sec:intro}

Substructures are ubiquitous in protoplanetary disks, particularly in the dust density distribution exhibited by high-angular-resolution observations \citep{Andrews2020,bae2023_PPPVII}. The Atacama Large Millimeter Array (ALMA) has revealed a variety of substructures, where a large population of rings and gaps are shown in continuum observations, and to a lesser extent, in molecular line emissions (e.g., \citealt{vanderMarel2019,zhang2021}).

Advances in spatial resolution have allowed us to resolve non-axisymmetric substructures within gaps, including systems such as PDS 70 \citep{Benisty2021}, HD 163296 \citep{Isella2018}, HD 100546 \citep{Perez2020}, HD 97048 \citep{Pinte2019}, and LkCa 15 \citep{Feng2022}. These substructures may emerge from the gravitational perturbations from embedded planets (e.g. \citealt{bae2023_PPPVII}), which vary in morphology depending on the underlying dynamics. Point-like emissions are generally associated with an accreting planet surrounded by a circumplanetary disk (CPD) \citep{Perez2015,Szulagyi2018}, while more azimuthally extended crescents or clumps may be related to the co-orbital stable Lagrange points $L_4$ and $L_5$ of a star-planet system \citep{Feng2022}. The particular disk in PDS 70 may host both: a CPD around planet c \citep{Benisty2021} and dust clump near L5 of planet b \citep{Balsalobre-Ruza2023}.

%High-resolution observations of protoplanetary disks in the dust continuum have revealed non-axisymmetric features that could potentially be attributed to Lagrange points. \cite{Long2022} detected two robust non-axisymmetric emission features at putative $L_4$ and $L_5$ points in the disk LkCa 15. Similarly, \cite{Balsalobre-Ruza2023} found a highly significant emission at the position of the $L_5$ region of the innermost planet in the disk PDS 70. 

Numerical simulations further support the notion that dust clumps and/or crescents inside\footnote{Crescents at the edges of the density gaps or in density rings may naturally arise due to the Rossby Wave Instability \citep{lovelace99}, possibly but not necessarily due to embedded  planets  (e.g., \citealt{de-val-borro07,chang23}).} density gaps are associated with the stable Lagrange points $L_4$ and $L_5$ of embedded planets \citep{Lyra2009,Montesinos2020}. These Lagrange islands are not only a region of librating streamlines, but also a region with higher pressure than its surroundings, thus naturally trapping dust. A fully worked example includes HD163296 where \citet{Rodenkirch2021} showed that the prominent crescent inside a dust gap could be naturally explained as material lingering near the $L_5$ point of a Jupiter-mass planet. This model was subsequently refined by \citet{Garrido-Deutelmoser2023} arguing for a pair of sub-Saturns near the 4:3 resonance to better explain the offset of the crescent and the shallow gas gaps derived from the CO emission \citep{Zhang2018}. Regardless of these details, the dynamics consistently invokes material clustered in the {\it trailing Lagrange point $L_5$}.

All in all, the handful observations (LkCa 15, HD 163296, PDS 70) and previous simulation studies all indicate that the clustering of dust takes place predominantly in the trailing Lagrange point $L_5$. The reason behind this asymmetry remains unaddressed and is the main question we tackle in this work. 

In the context of restricted three-body problem (RTBP) the librating regions around $L_4$ and $L_5$ are symmetric (e.g., \citealt{MD2000}). A natural way to break the symmetry is to have a net radial drift between the test particles and the planet: $L_4$ ($L_5$) traps more particles for converging (diverging) radial drifts, for instance, as a result of inward (outward) planet migration \citep[e.g.,][]{Murray1994, Masset2002, Sicardy2003}. 
%\cp{(cite more paper Masset ($\alpha_{ss}$), Murray)}
However, hydrodynamic simulations show a strong asymmetry in the gas even in the absence of a radial drift (see, e.g., Figure 6 in \citealt{garrido2022} for a non-migrating planet in a low-viscosity disk).

In this work, we study the gas dynamics near the co-orbital region of a non-migrating planet in order to explain the origin of the asymmetry between $L_4$ and $L_5$. By linking the level of asymmetry, or lack thereof, to a physical process, we aim to constrain the disk properties and the mass of embedded planets.

\section{Setup} \label{sec:setup}

%In our analysis, 
We consider a planet of mass $m_{\rm p}$, embedded in a two-dimensional gaseous disk, moving in a fixed circular orbit of semi-major axis $r_{\rm p}$ around a central star of mass $M_{\star}$. To simulate the system we use the hydrodynamic code FARGO3D\footnote{Publicly available at \href{https://github.com/FARGO3D/fargo3d}{github.com/FARGO3D/fargo3d}} \citep{Masset2000, Benitez2016}. %which solves the Navier-Stokes equations accounting for two-dimensional quantities. 
%In this set of equations, 
%These equations are the conservation of mass
%\begin{equation}
%    \frac{\partial \Sigma}{\partial t} + \nabla \cdot (\Sigma \textbf{v})=0,
%\label{eq:cons_mass}
%\end{equation}

%\noindent
%and the conservation of momentum
%\begin{equation}
%    \Sigma \left( \frac{\partial \textbf{v}}{\partial t}+\textbf{v} \cdot \nabla \textbf{v} \right) = -\nabla P -\Sigma\nabla{\Phi} + \nabla \cdot \vec{T},
%\label{eq:cons_momentum}
%\end{equation} 

%\noindent
%where $\Sigma$ is the surface density, $\textbf{v}$ the flow velocity vector, $\Phi$ include the gravitational potential of the star, planet and indirect terms, $\vec{T}$ the viscous stress tensor, and $P$ the pressure. 

For the initial surface density of the disk, we adopt a power-law profile with an exponent denoted by $\alpha$
\begin{equation}
  \Sigma(r) = \Sigma_{\rm p}\left(\frac{r}{r_{\rm p}}\right)^{\alpha},
  \label{ec:sigma0}
\end{equation}
with $\Sigma_{\rm p}=\Sigma(r_{\rm p})$. Regarding the gas pressure, we assume a locally isothermal equation of state $P = \Sigma c_{\rm s}^{2}$, where the sound speed, $c_{\rm s}$, is defined as
\begin{equation}
  c_{\rm s}(r) = h v_{\rm K} = c_{\rm s,p} \left(\frac{r}{r_{\rm p}}\right)^{2\beta},
\end{equation}
being $v_{\rm K}=\sqrt{GM_\star/r}$ the Keplerian velocity, $c_{\rm s,p}=h_{\rm p} v_{\rm K,p}$ the sound speed at the planet's orbital radius, and $h$ the disk aspect ratio, given by
%, which is also described by a power-law with an exponent $\beta$
\begin{equation}
    h = h_{\rm p}\left(\frac{r}{r_{\rm p}}\right)^{f}\,,
\end{equation}
with $f$ the flaring index and $h_{\rm p}=h(r_{\rm p})$ the disk aspect ratio at the planet's radius. %Both quantities are related through $H=c_{\rm s}/\Omega$, with the orbital frequency $\Omega$ deviating from the pure Keplerian value $\Omega_{\rm K}^2=GM_{\star}/r^{3}$ due to the pressure gradient
%\begin{equation}
%    \Omega^2=\Omega_{\rm K}^2+\frac{1}{r\Sigma}\frac{dP}{dr}.
%\end{equation}
%The disk's scale height can be expressed in terms of the aspect ratio as
%\begin{equation}
%   \frac{H}{r}=h(r)=h_{\rm p}\left(\frac{r}{r_{\rm p}}\right)^f,
%\end{equation}
The flaring index is related to the sound speed exponent by $2\beta=f-1/2$, and also to the temperature profile of the disk through $T\propto r^{\beta}$.

In this work, we adopt the $\alpha$-viscosity prescription, in which the kinematic viscosity of the gas is modeled as $\nu=\alpha_{\rm ss}c_{\rm s} H$, with $H=hr$ the disk scale height and $\alpha_{\rm ss}$ a constant dimensionless parameter \citep{Shakura1973}.

The planet's gravitational potential is modeled as %of the planet,
\begin{equation}
    \Phi_{\rm p}(\textbf{r})=-\frac{Gm_{\rm p}}{\sqrt{|\textbf{r}-\textbf{r}_{\rm p}|+\epsilon}}+\frac{Gm_{\rm p}\textbf{r}\cdot\textbf{r}_{\rm p}}{r_{\rm p}^3}\,,
\end{equation}
allowing us to have analogues of the Lagrange points in the restricted three-body problem (RTBP). In the hydrodynamic framework, these correspond to the stagnation points, henceforth referred to as Lagrange points. The modified direct term in this equation includes the smoothing length $\epsilon$, which accounts for the vertical extent of the disk in our two-dimensional setup \citep[see e.g.,][]{Muller2012}. In this work, we set $\epsilon=0.6H_{\rm p}$ \footnote{Variations in the smoothing length $\epsilon$ alters the amount of excited azimuthal modes of the planet-induced density perturbations, which can lead to non-realistic outcomes (see e.g., \citealt{Ogilvie+2002, Muller2012})}.%, with $H_{\rm p}$ the disk's scale height at the planet's orbital radius.%, as this is the value commonly used in  two-dimensional simulations.

Lastly, the mass of the disk is set to be sufficiently small to ensure that it remains gravitationally stable, and that the planet does not migrate.

\begin{figure*}[t!]
\begin{center}
    \includegraphics[width=\linewidth]{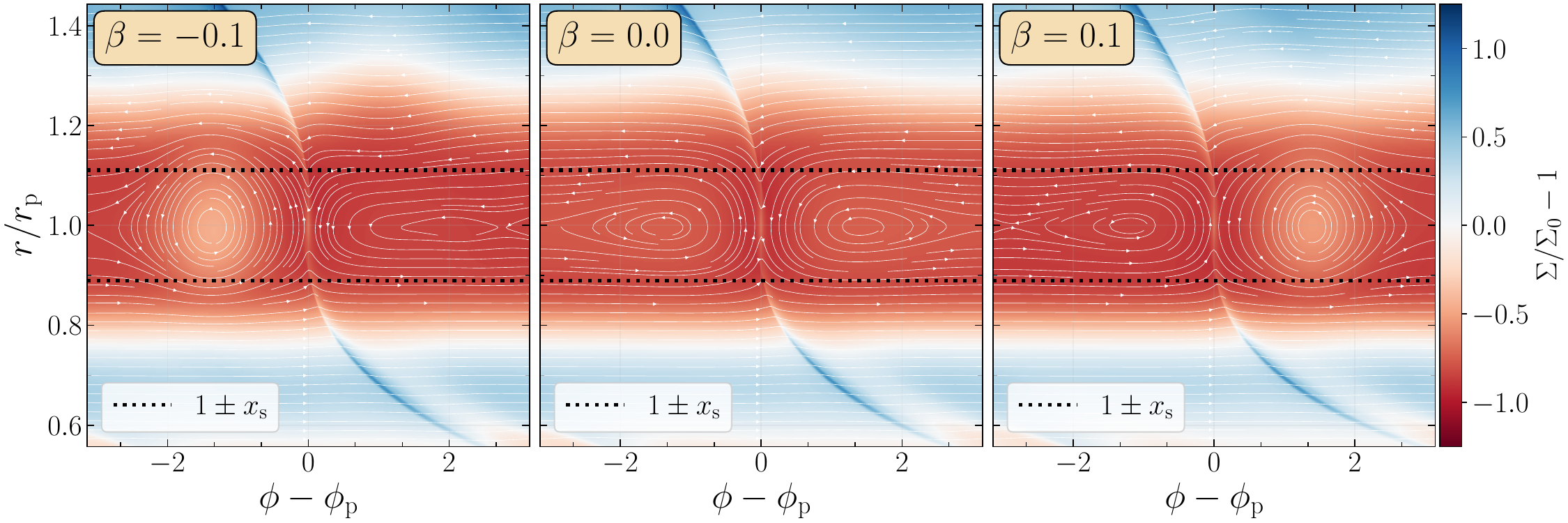}
\end{center}
    \caption{Two-dimensional maps of the normalized gas surface density $\Sigma/\Sigma_0 - 1$ after $\sim1500$ orbits, with $\Sigma_0$ the initial density profile given by equation \ref{ec:sigma0}, for three different values of $\beta$, and with the fiducial parameters for $m_{\rm p}$, $h_{\rm p}$ and $\alpha$. The planet is located at $r/r_{\rm p}=1$ and $\phi=0$. The black dashed horizontal lines correspond to the horse-shoe semi-width $x_{\rm s}$ (see Section \ref{sec:3.1.}).}
    \label{fig:asymmetry}
\end{figure*}

\subsection{Fiducial model} \label{sec:2.1.}

Our fiducial system consists of a central solar-mass star and a planet with $m_{\rm p}=125$ $M_{\oplus}$ embedded in a globally isothermal disk ($\beta=0$), which extends from $r_{\rm in}/r_{\rm p} = 0.3$ to $r_{\rm out}/r_{\rm p} = 3.0$. At the radius of the planet, the disk has an aspect ratio $h_{\rm p}=0.07$. We also set a low viscosity environment with $\alpha_{\rm ss}=10^{-4}$.

%This implies a disk mass of approximately one Earth mass for the initial surface density profile $\Sigma(r)=1.7$ $(r/r_p)^{-2}$ gr/cm$^2$. This represents $0.1$\% of the minimum mass solar nebula \citep{Hayashi1981}.

The computational grid needed to solve the hydrodynamics is composed of 512 cells logarithmically spaced along the radial direction within $[r_{\rm in}/r_{\rm p},r_{\rm out}/r_{\rm p}]$, and 1024 cells uniformly spaced azimuthally within the $[0,2\pi]$ domain.

\subsection{Simulation set} \label{sec:2.2.}

To characterize the distribution of gas around $L_4$ and $L_5$ we conduct a series of simulations varying the following parameters; the gas surface density exponent $\alpha$, the temperature exponent $\beta$, the mass of the planet $m_{\rm p}$, and the aspect ratio of the disk $h_{\rm p}$.

For $m_{\rm p}$ and $h_{\rm p}$, we use a uniform $15\times15$ grid covering planetary masses from 30 to 300 $M_{\oplus}$ and aspect ratios from 1/30 to 2/15, selecting only the systems with planets opening gaps between $0.2<\Sigma_{\rm gap}/\Sigma_0<0.02$. This results in 83 simulations, all carried out with fixed values of $\beta = -0.1$ and $\alpha = 1.6$. In addition, we run our fiducial model for four different temperature gradients: $\pm0.05$ and $\pm0.1$, using $\alpha = 2(2\beta + 1)$ in each case. Finally, we explore the fiducial model for three values of the gas surface density exponent: 0.5, 1.0, and 1.5, all with $\beta = -0.1$.

%\section{Lagrange Points Asymmetry: dependence on $\beta$} \label{sec:beta_dependence}
\section{\texorpdfstring{Lagrange Points Asymmetry: dependence on $\beta$}{}} \label{sec:beta_dependence}
%To determine which parameters are causing the distribution of gas in the Lagrange points we performed an extensive exploration of the temperature and gas surface density gradients, $\beta$ and $\alpha$ respectively.

%After running the set of simulations previously defined,

By analyzing our suite of simulations, we find that the distribution of gas around $L_4$ and $L_5$ is solely regulated by the temperature gradient in the disk. Furthermore, this distribution is found to be very robust to variations in $\alpha$, $m_{\rm p}$ and $h_{\rm p}$. 

An example is shown in Figure \ref{fig:asymmetry}, where a globally isothermal disk ($\beta=0$) displays a symmetrical distribution of gas between $L_4$ and $L_5$. Additionally, it depicts how this symmetry is disrupted by changes to the sign of the temperature gradient. If the temperature increases with radius ($\beta>0$) $L_4$ retains more gas than $L_5$ (see the right panel in Figure \ref{fig:asymmetry}). In this case, the libration region around the $L_4$ point expands radially compared to the symmetric case, while the one around $L_5$ does not vary significantly. 

Conversely, if the temperature radially decreases $L_5$ preserves more material than $L_4$ (see the left panel in Figure \ref{fig:asymmetry}). Here, the gas rotating around $L_5$ expands radially, with no change in the $L_4$ region with respect to the symmetric case.

Furthermore, for both $\beta=-0.1$ and $\beta=0.1$ the peaks in the normalized gas surface density $\Sigma/\Sigma_0 - 1$ around the most prominent Lagrange point after 1500 orbits are very similar, with absolute values of $\sim0.45$ and $\sim0.53$, respectively.

%{\color{red} I would add more description of the figure. For example, describe not only the amount of mass but also the shape characterized through the streamlines. It would be relevant to mark the Lagrange Points in Fig. 1 and use the marks to explain their features in this section. Finally, quantify/mention the density contrast in each case.}

\subsection{Time evolution} \label{sec:3.1.}

%Here, we measure the time evolution of the gas density contrast between $L_4$ and $L_5$. 

Having depicted the morphology of the structures around $L_4$ and $L_5$ at a specific time in Figure \ref{fig:asymmetry}, we now measure the time evolution of the gas density contrast between them.

Previous work, such as \cite{Garrido-Deutelmoser2023, Montesinos2020}, estimated this evolution using boxes centered on the RTBP locations of each point weighting them by the evolution of the unstable $L_3$ point. However, since there is an azimuthal offset in the position of $L_4$ and $L_5$ with respect to the RTBP values (see Section \ref{sec:offset}), and given the non-isolated evolution of $L_3$, we use a different method to measure the evolution.

We first split in two the azimuthal domain at the angular position of the planet ($\phi_{\rm p}=0$). In the co-rotating frame, the first region goes from $-\pi$ to 0 and the second one from 0 to $\pi$. Then, we subtract the gas surface density of these two regions normalized by the azimuthally-averaged gas surface density and measure the contrast between $L_4$ and $L_5$, since each region encloses only one of the point. Finally, we compute the absolute maximum within a rectangular box for this contrasted region.

\begin{figure}[t!]
\centering
\includegraphics[width=1\linewidth]{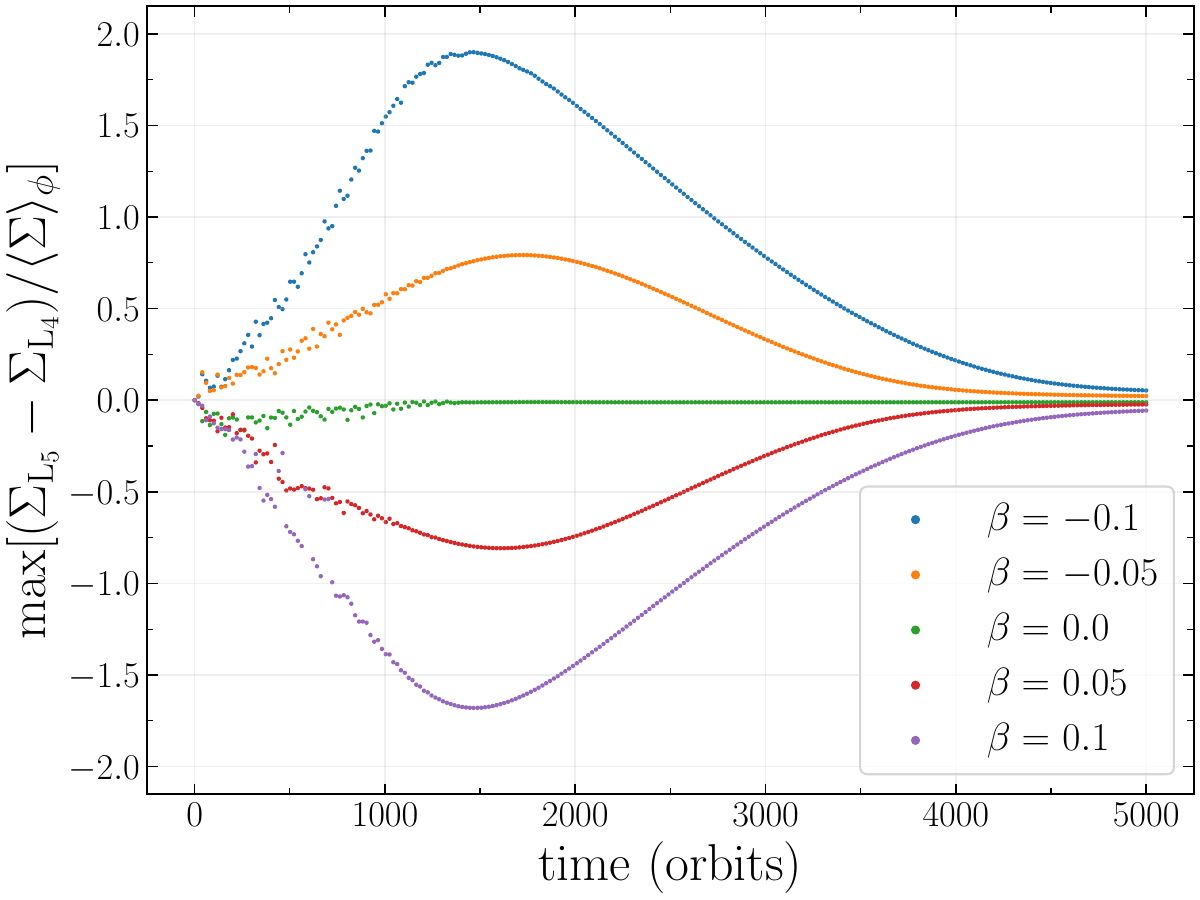}
\caption{Time evolution of the gas density contrast between $L_4$ and $L_5$ for different temperature gradients $\beta$ displayed as the maximum contrast between these regions. All simulations have the fiducial parameters $m_{\rm p}=125\text{ }M_\oplus$, $h_{\rm p}=0.07$, and $\alpha_{\rm ss}=10^{-4}$.}
\label{fig:dens_evo}
%\vspace{-0.3cm}
\end{figure}

The azimuthal extent of the box is $\pi/4$, as in \citet{Garrido-Deutelmoser2023}, and it is centered at the angular location of the Lagrange points (see Section \ref{sec:offset}).

For the radial range we use the semi-width of the horseshoe region $x_{\rm s}$, which corresponds to the outermost horseshoe orbit. As a proxy of $x_{\rm s}$ in our simulations, we use the quantity defined by \cite{Jimenez2017}
\begin{equation}
    x_{\rm s}=H_{\rm p}\left(\frac{1.05Q^{1/2}+3.4Q^{7/3}}{1+2Q^2}\right),
    \label{ec:xs}
\end{equation}
which is valid for a wide range of masses in 3D simulations (i.e., with negligible smoothing length). In this expression, $Q$ is the ratio between the mass of the planet and the so-called thermal mass, which is defined as 
\begin{equation}
    M_{\rm th}\approx M_{\star}h_{\rm p}^3,
    \label{ec:thermal_mass}
\end{equation}
%\jg{By this definition of thermal mass, should be a 3 in every $Q$ of equation \ref{ec:xs}, this is how the horizontal lines in top figure \ref{fig:asymmetry} have been calculated.}
and corresponds to the mass at which the Hill radius of a planet equals the pressure scale height of the disk (off by a factor of 3). 

%\cc{This parameter is often used as a proxy to characterize the transition between the planet exciting linear spiral density waves ($m_p\ll M_{\rm th}$) and non-linear waves  ($m_p\gtrsim M_{\rm th}$).} 

%The advantage of taking the absolute maxima within the radial ring $|r-r_{\rm p}|\leq x_{\rm s}$ is that we are more sensitive to differences in the gas surface density produced by low-mass planets.

%With this metric we measure how the evolution of the gas surface density changes for different values of $\beta$, with the other parameters fixed.

Figure \ref{fig:dens_evo} shows the time evolution of the gas distribution for different values of $\beta$. In the globally isothermal case ($\beta=0$), the gas structures around both $L_4$ and $L_5$ evolve in a nearly identical manner, so the time evolution for the contrast between them is a horizontal line.

However, when a temperature gradient is introduced into the disk ($\beta\neq0$), the evolution becomes non-uniform. After an initial phase of approximately linear growth, the gas density reaches a maximum and subsequently decays to zero as material is gradually depleted from the co-orbital region. The maximum contrast between $L_4$ and $L_5$ increases monotonically with $\beta$; in other words, steeper temperature gradients lead to larger asymmetries. Furthermore, for the same absolute value of $\beta$, the time evolution of the contrast is very similar (see for example the red and orange curves in Figure \ref{fig:dens_evo}).

%{\color{red} Again, more description would benefit the paper. You should explain what happens for different values of $\beta$. So far it is not clear how you are locating the Lagrange points. I would add explicitly that their are actual stagnation points in your simulation. This is only mentioned in the setup}

\begin{figure}[t!]
    \centering
    \includegraphics[width=1\linewidth]{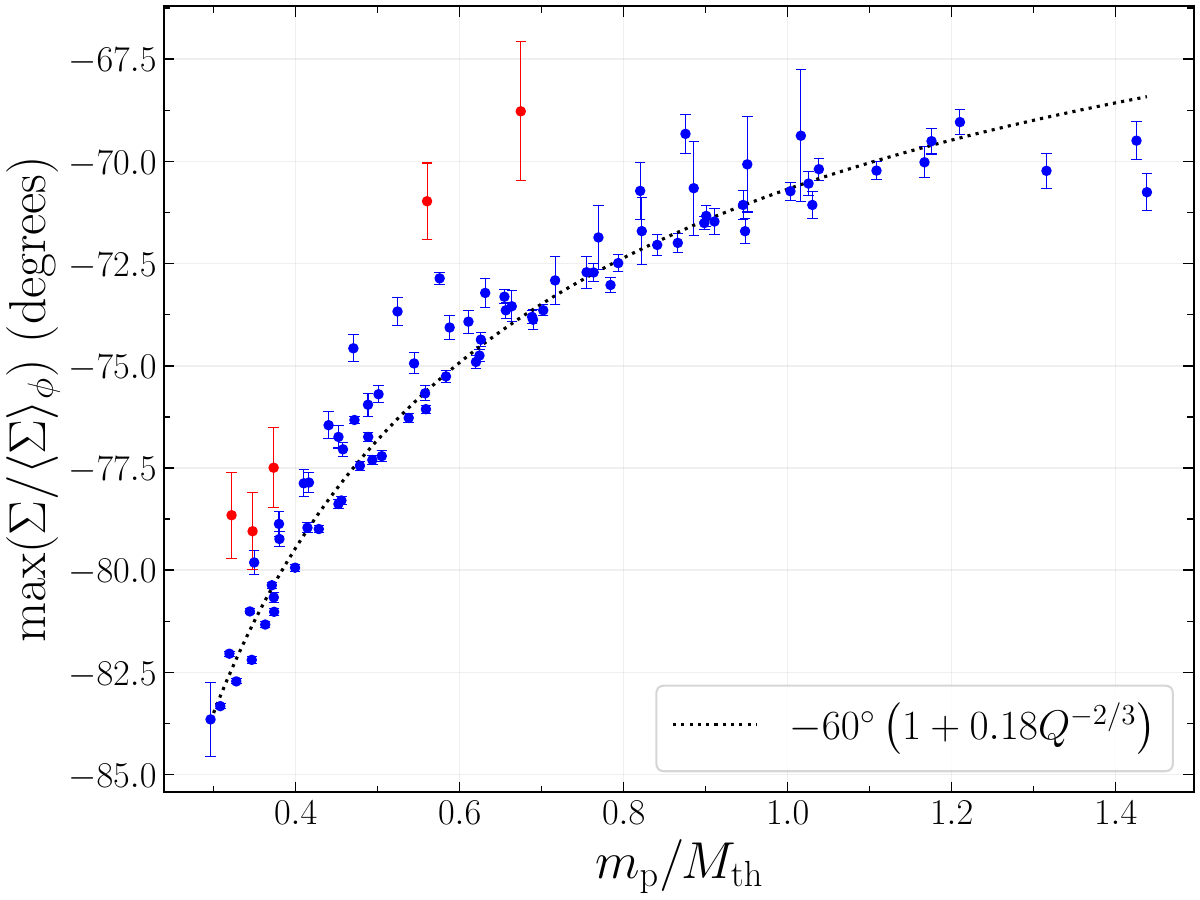}
    \caption{Azimuthal location of the gas surface density peak as a function of $Q=m_{\rm p}/M_{\rm th}$. Each point displays the median with an error bar from the 5th to 95th percentiles computed from an uniformly time-sampled window of $1000-5000$ orbits. The black dashed line is the best power-law fit computed using the median values (Equation \ref{ec:offset}). The red dots deviate significantly from the black dashed line and have large errors. These outliers correspond to systems with $m_{\rm p}\gtrsim185$ $M_{\oplus}$, which have a more chaotic evolution than those below this mass (see Section \ref{sec:h_and_m}).}
    \label{fig:offset}
\end{figure}

\subsection{Azimuthal location of the Lagrange points} \label{sec:offset}

Our simulations show that $L_4$ and $L_5$ do not remain fixed at $\pm60^{\circ}$ from the planet, as predicted by the RTBP, but instead shift in azimuth depending on the planet’s mass and the disk’s aspect ratio.

%Figure 4
\begin{figure*}[t!]
    \includegraphics[width=\linewidth]{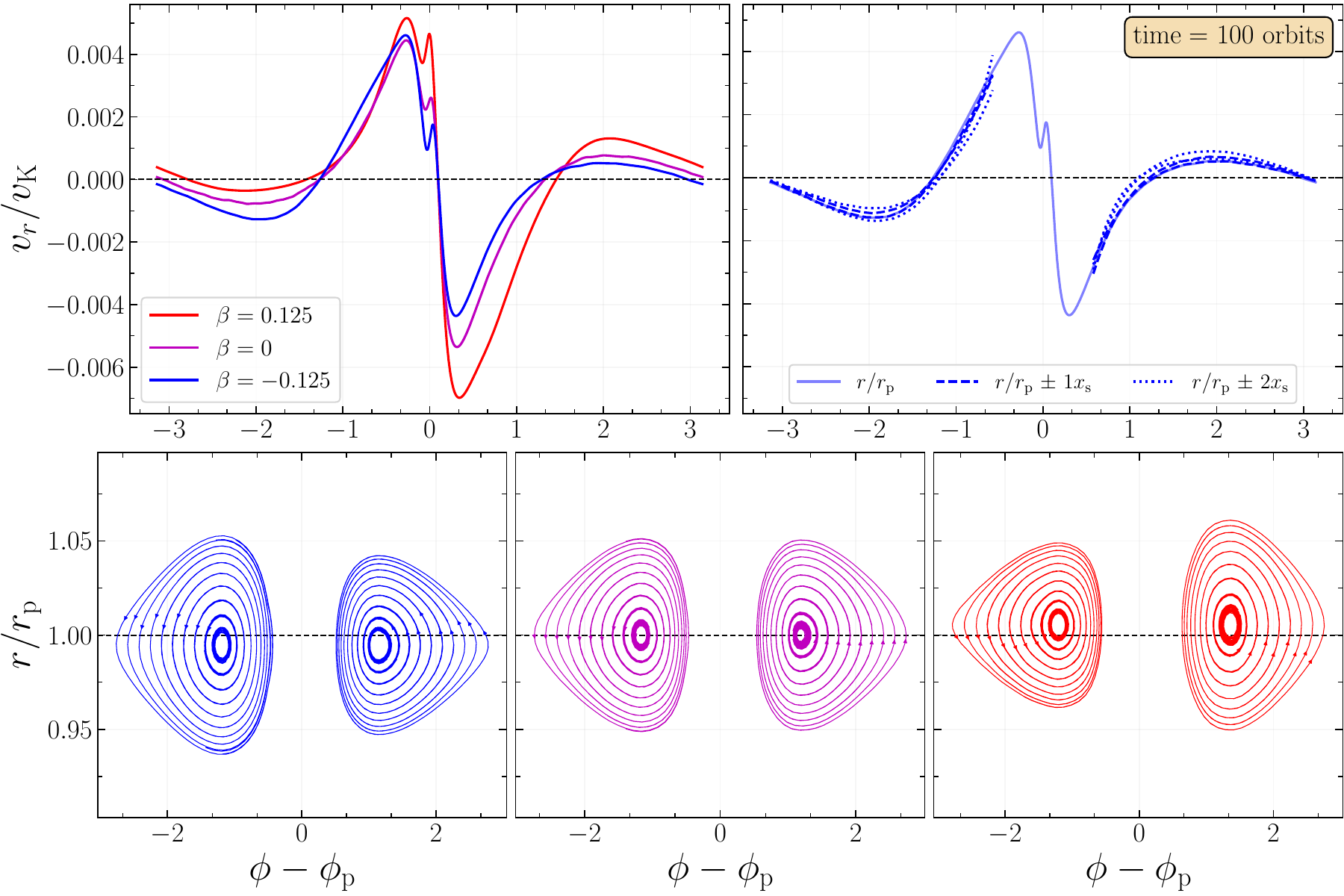}
    \caption{\textit{Top Left:} Radial velocity curve at the co-rotation radius $r=r_{\rm p}$ for a planet with $m_{\rm p}=200$ $M_{\oplus}$, and a disk with an aspect ratio of $h_{\rm p}=0.13$. The red curve correspond to a system with $\beta=0.125$, while the blue curve is a system with $\beta=-0.125$. \textit{Top Right:} Radial cuts to the radial velocity field for $\beta=-0.125$ at different horseshoe semi-widths with respect to the co-rotation radius $r=r_{\rm p}$. We masked the region $|\phi-\phi_{\rm p}|<0.7$ for the radial velocity cuts outside co-rotation. \textit{Bottom:} Streamlines for the librating material computed with the semi-analytical model described in Section \ref{sec:model}, for different values of the temperature gradient. The blue and red lines have the same $\beta$ as the curves in the top left panel, while the magenta lines have $\beta=0$. By plugging the radial velocity profiles from the top left panel into our analytical model, the symmetric purple streamlines deform asymmetrically; expanding the region around $L_5$ for $\beta=-0.125$ and expanding the region around $L_4$ for $\beta=0.125$, resembling the trend shown in Figure \ref{fig:asymmetry}.}
    \label{fig:vr_effect}
\end{figure*}

Since we are interested in tracing the gas structures in the co-orbital region, we measure the offset of the peaks in the azimuthally-averaged gas surface density profile---these are the locations around which the gas accumulates---instead of the stagnation points. Nonetheless, the stagnation points are very close to the locations of the peaks.

Due to the scatter observed throughout the time evolution of most simulations, we use the median as a measure of the azimuthal location of the peak near the Lagrange  points.

%It is worth while to mention that we only compute this quantity for the $L_4$ point as the offset is approximately equal for both $L_4$ and $L_5$, and also because for high contrast systems is almost impossible to even identify the $L_5$ point.

Figure \ref{fig:offset} shows that the angular offset in the azimuthal position of the peak is non-negligible, varying from a few degrees up to 25$^{\circ}$ in the more extreme cases. We also observe that this angular offset increases monotonically with $Q$, a trend that can be approximately described by the fitting formula
\begin{equation}
    |\phi_{\rm max}|\approx60^{\circ}\left(1+0.18Q^{-2/3}\right),
    \label{ec:offset}
\end{equation}
where $\phi_{\rm max}$ is the angular position of the peak, and $Q$ is the ratio between the mass the planet and the thermal mass. 

%\jg{Check if this definition of $Q$ correspond with the thermal mass including the factor $3$}

However, we find that some systems deviate significantly from this trend (see the red dots in Figure \ref{fig:dens_evo}). These outliers correspond to planetary masses above 240 $M_{\oplus}$, which are cases with a more chaotic evolution driven by strong vortices induced by the Rossby Wave Instability (RWI) at the edges of the gaps opened by these very massive planets (see section \ref{sec:h_and_m}).

%Figure 5
\begin{figure*}[t!]
    \includegraphics[width=\linewidth]{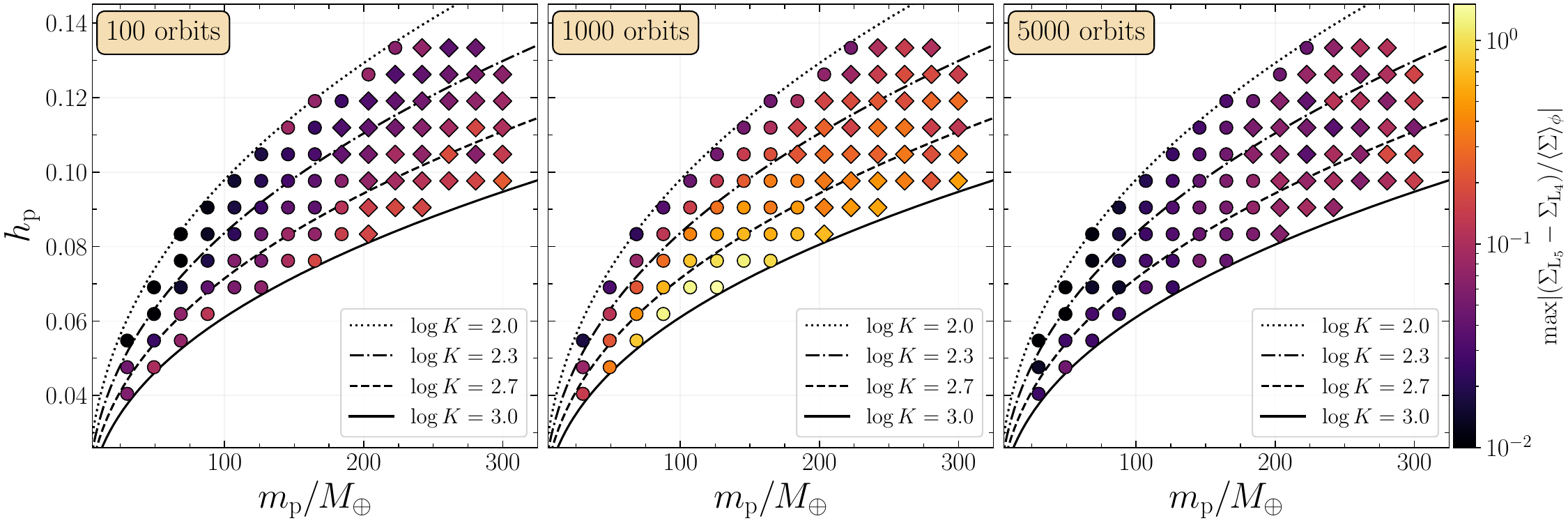}
    \caption{Gas distribution around $L_4$ and $L_5$ for all simulations computed with the method described in Section \ref{sec:3.1.} and shown in Figure \ref{fig:dens_evo}. From left to right, the panels depict the contrast in gas between the Lagrange points at 100, 1000, and 5000 orbits. The circles are simulations with planetary masses below 185 $M_{\oplus}$, while the diamonds represent the ones with masses above 185 $M_{\oplus}$. The various black lines correspond to different gap depths given by the $K$ parameter (see the legend).}
    \label{fig:sim_grid}
\end{figure*}

We attribute these deviations to the pressure gradient in the co-orbital region. When this region is dominated by gas pressure (i.e., $Q\ll1$), the second term in the right-hand side of equation \ref{ec:offset} increases, and so does the angular offset. Conversely, if the planet dominates its vicinity (i.e., $Q\gg1$), this term becomes negligible, resembling the result of the RTBP (see Figure \ref{fig:offset}).

\begin{comment}
\begin{figure*}[t!]
\begin{center}
    \includegraphics[width=\linewidth]{Figs/toy_model.pdf}
\end{center}
    \caption{Spatial distribution of $\sim6700$ test particles at 0 and 20000 orbits. The planet has a mass of $150$ $M_{\oplus}$, and the pressure correction is characterized by $\beta=-0.5$ and $h_{\rm p}=0.1$. The colors indicate the initial location of the particles around each of the three co-orbital Lagrange points, $L_3$ (red), $L_4$ (green) and $L_5$ (blue). The particles are uniformly distributed around these points. The black dashed lines are the energy levels derived from the RTBP.}
    \label{fig:toy_model}
\end{figure*}
\end{comment}

\section{Lagrange Points Asymmetry: Semi-Analytical Model} \label{sec:model}

%\cp{CP: I suggest to shorten and write more linear. First say that you don't see mass transfer. Second, introduce OL06, write the equations for the velocities with $\delta v_r(\phi)$ and $\delta v_\phi$, say the former makes the treat. Remember that a paper is not a report.}

%Based on the approximation from \cite{Ogilvie2006}, we constructed a simple analytical model that describes the gas dynamics in the co-orbital region (see Appendix \ref{A:SA_model} for details). Although it displays some of the features we observe in the hydrodynamic simulations performed with FARGO3D, we concluded that the co-orbital gas dynamics are not fully captured by this model, as some outcomes of disk-planet interactions are not included in this model.

%Thus, we decided to use a passive tracer instead to describe the co-orbital gas dynamics. To accomplish that, we first selected an ideal system where the evolution of the gas around the Lagrange point is smooth, well-defined and with a significant contrast. Then, we considered a window of 1000 orbits centered at the peak of the evolution. To trace the gas dynamics within this window, we utilized a well-coupled dust species with $\rm{St}\ll1$. The dust was then initialized as two bi-dimensional Gaussian profiles centered at $L_4$ and $L_5$. For the characteristic widths of the profiles we used $\sigma_{\phi}=\pi/8$ and $\sigma_r=0.4x_{\rm s}$ . Finally, we re-ran the system within the 1000 orbits window with this well-coupled dust species and with no feedback to the gas. 

By using a well-coupled dust species (i.e., $\rm{St}\ll1$) to trace the co-orbital gas we noticed that there was no significant mass transfer from the region surrounding one point to the other. However, we observed that as one of the Lagrange points became more prominent than the other, it expanded in both radius and azimuth significantly increasing its area relative to the other, which is consistent with the streamlines in Figure \ref{fig:asymmetry}.

To explore this process further, we analyzed the velocity field of the gas in the early stages of its evolution combined with the analytical model proposed by \citealt{Ogilvie2006}, which approximates the co-orbital velocity field as
\begin{flalign}
    \frac{v_{\phi}}{r_{\rm p}\Omega_{\rm p}}&=\frac{m_{\rm p}}{M_{\star}}\frac{x}{2}\left(s^2+x^2\right)^{-3/2}-\frac{3}{2}x+\frac{h_{\rm p}^2\beta}{2},\\
    \frac{v_r}{r_{\rm p}\Omega_{\rm p}}&=\frac{m_{\rm p}}{M_{\star}}s\sqrt{4-s^2}\left[1-\left(s^2+x^2\right)^{-3/2}\right]+\delta v_r,
    \label{ec:OL06}
\end{flalign}
where $x=r/r_{\rm p}-1$ is the distance with respect to co-rotation and $s=2\sin(\phi/2)$. 

The first term on the right-hand side of both equations corresponds to the perturbation of the planet. The last two terms in the azimuthal velocity equation are the Keplerian motion and the pressure correction---this correction produces the vertical shift observed in the streamlines of the bottom panels in Figure \ref{fig:vr_effect}, depending on the sign of $\beta$---respectively. The last term in the radial velocity equation is a subtle yet non-negligible radial velocity profile that is established in the protoplanetary disk due to the disk-planet interaction.

This small radial velocity component $\delta v_r$ changes as a function of the temperature gradient of the disk, as shown in the upper left panel in Figure \ref{fig:vr_effect}. Furthermore, it does not vary if we move out of co-rotation, as observed in the upper right panel in Figure \ref{fig:vr_effect}. 

By taking this $\delta v_r$ component from our simulations we can effectively deform the libration region as a function of the temperature gradient $\beta$, reproducing the trend shown in Figure \ref{fig:asymmetry}.

Near the point favored by the temperature gradient of the disk, there is a steeper azimuthal gradient in the radial velocity compared to the other point (see the top left panel in Figure \ref{fig:vr_effect}). That is, the librating material around this location has a larger radial amplitude induced by the larger radial velocity. As a result, the librating region around this Lagrange point expands, and thus it can retain more material than the other point, creating an asymmetry (see bottom panels in Figure \ref{fig:vr_effect} for $\beta\neq0$). As the radial velocity profile keeps evolving in time, the co-orbital gas develops the trend depicted in Figure \ref{fig:dens_evo} for the gas contrast between $L_4$ and $L_5$. 

%This initial coupling between the evolution of the radial velocity field and the co-orbital gas could be modeled using linear theory of disk-planet interaction (e.g., \citealt{Miranda+2019}). In this regime, we would expect the radial velocity field to depend on the temperature gradient of the disk $\beta$, producing the observed velocity gradients at the Lagrange points $L_4$ and $L_5$.

\section{Parameter Space Exploration} \label{sec:parameters}

Given the dependence we have found between the gas distribution and the temperature gradient in the disk, we now explore how this phenomenon changes its amplitude and longevity with variations in $\alpha$, $m_{\rm p}$ and $h_{\rm p}$ for a fixed value of $\beta$.

%Figure 6
\begin{figure*}[hbt!]
    \includegraphics[width=\linewidth]{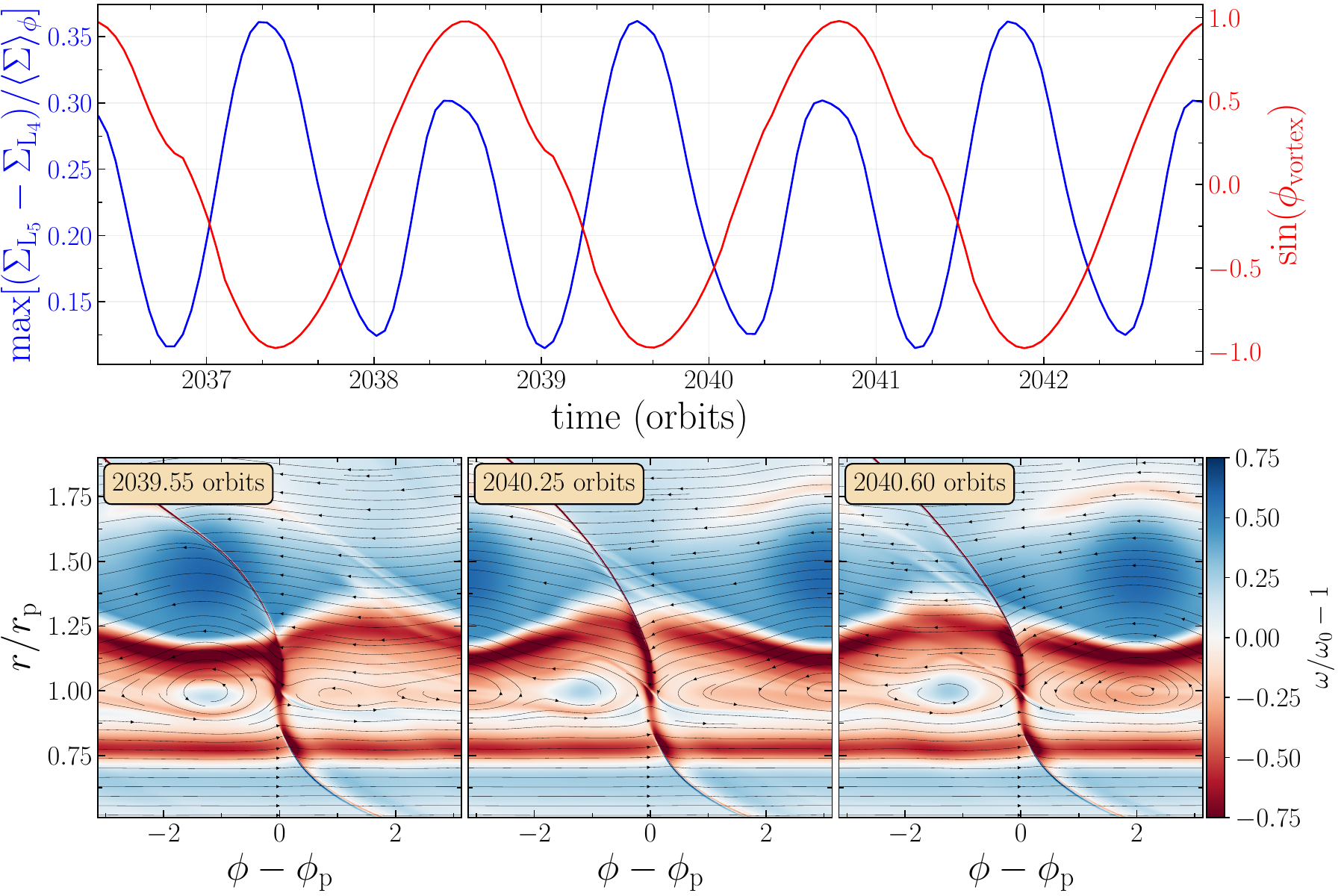}
    \caption{\textit{Top:} The blue curve shows the time evolution of a simulation with $m_{\rm p}>185$ $M_{\oplus}$, over 2037–2042 orbits. The red curve represents the evolution of the azimuthal location of the vortex formed outside the outer edge of the gap. \textit{Bottom:} From left to right, vorticity maps illustrate three different positions of the vortex (blue blob at $\sim1.5$ au). The outer panels correspond to conjunctions with $L_5$ and $L_4$, respectively, aligning with the two different peaks in the blue curve. When the vortex is located at $L_3$ (middle panel) or at the position of the planet, the blue curve reaches its minima.}
    \label{fig:hm_regime}
\end{figure*}

Figure \ref{fig:sim_grid} depicts the maximum gas surface density contrast between $L_4$ and $L_5$ for our suite of simulations as a function of planetary mass and disk aspect ratio. We show this at three different times: 100, 1000 and 5000 orbits. At 100 orbits, one can observe the onset of the asymmetry as the planet starts carving a gap. Then, at 1000 orbits, one sees the approximate time around which the gas contrast between the Lagrange points peaks for most cases. Finally, after 5000 orbits the contrast decays to zero in most systems, with values similar to those observed at 100 orbits. 

%After a protoplanet is formed it rapidly interacts and perturbs its gaseous surroundings. For the sake of this study we will not take into account the planet formation phase, as it is still a major unknown in the field (references with hypothesis of early-stages planet formation).

As we shall show, a key physical parameter in this analysis is the morphology of the gap opened by a planet. For a wide range of planetary masses, if the gap is too shallow, then the contrast produced between $L_4$ and $L_5$ is negligible. However, if the perturbation of the planet is strong enough it can lead to the formation of a deeper gap, enhancing the contrasts.

%According to spiral wave theory (e.g. \citet{GR80}), at the locations of Lindblad resonances, different harmonics of surface density perturbations propagate away from the planet. These waves transport angular momentum redistributing gas within the disk. 

The dependence of this gap-opening process on the system parameters has been derived empirically by \cite{Kanagawa2017}. They showed that the steady-state gap morphology is directly related to
\begin{equation}
    K = \left(\frac{m_{\rm p}}{M_{\star}}\right)^2\left(\frac{H_{\rm p}}{r_{\rm p}}\right)^{-5}\alpha_{\rm ss}^{-1},
\label{ec:K}
\end{equation}

\noindent
which predicts both the width and the depth of a gap, with wider and deeper gaps for larger values of $K$. 

We find that for $2.0<\log K<3.0$ the gas distribution can lead to a significant contrast between $L_4$ and $L_5$, below $2.0$ the gaps are too shallow and the contrast is negligible, and above $3.0$ the gap is too deep so the process occurs on a very short time scale and becomes highly chaotic. Furthermore, in this gap regime we observe that the peak in the time evolution of the gas distribution is reached in most cases between 500 and 1000 orbits.

\subsection{Disk aspect ratio and planetary mass} \label{sec:h_and_m}

From the middle panel of figure \ref{fig:sim_grid} we observe that for a fixed planetary mass, colder disks produce higher gas contrasts between $L_4$ and $L_5$, while for a fixed aspect ratio the systems with more massive planets produce larger gas contrasts between $L_4$ and $L_5$.

Furthermore, there is a critical planetary mass that defines two different regimes for the time evolution of the gas around the Lagrange points. After visual inspection, we find that for systems with $m_{\rm p}\lesssim185$ $M_{\oplus}$ the time evolution is similar to that shown in Figure \ref{fig:dens_evo}, with a well-defined peak and subsequent asymptotic behavior towards zero. 

However, for planets above this threshold, the evolution becomes more chaotic. In this regime, overdensities form in the outer gap (see the prominent blue blob in the bottom panels of Figure \ref{fig:hm_regime}). As these structures circulate they interact with and perturb the co-orbital gas. For the most massive planets, when these features become very pronounced, the gas around $L_4$ and $L_5$ experiences strong perturbations, leading to oscillations in its time evolution. 

As shown in Figure \ref{fig:hm_regime}, when this vortex azimuthally aligns with $L_4$ and $L_5$, it induces the strongest perturbation in the co-orbital gas, resulting in the two repeating peaks observed in the gas contrast evolution for these systems. The period of these oscillations, for both the red and blue curves in Figure \ref{fig:hm_regime}, matches the synodic period of the vortex, which is located at approximately 0.5 au relative to the planet.

\subsection{Gas surface density profile and viscous accretion}

Although variations in the surface density profile of the disk do not change the gas distribution around $L_4$ and $L_5$, the maximum contrast achievable by a particular system depends on the relation between $\alpha$ and $\beta$.

In particular, when $\alpha\neq2(2\beta+1)$, there is a non-zero radial accretion of gas towards the central star\footnote{For accretion disks, the steady state equilibrium implies that the gradients of the gas surface density and the sound speed must satisfy the relation $\alpha=2(2\beta+1)$. For non-accreting disks, where $v_r\approx0$, this relation is no longer valid and both $\alpha$ and $\beta$ are free parameters in the system.}, which accelerates the depletion of the co-orbital material. However, since the simulations have low viscosity ($\alpha_{\rm ss}=10^{-4}$), this effect is subtle and can be effectively mitigated by setting $\alpha=2(2\beta+1)$.

Overall, we find that the small variations in the observed peaks within the simulations, caused by viscous accretion, are not significant and therefore do not change our conclusions.

\section{Summary and Discussion} \label{sec:summ_disc}

We have found that gaseous disks with radial temperature gradients inherently exhibit asymmetrical gas distributions around the Lagrange points $L_4$ and $L_5$. In particular, when $\beta>0$, $L_4$ retains more gas than $L_5$, while $\beta<0$ causes $L_5$ to preserve more material.

When $\beta$ is fixed and non-zero, variations in the mass of the embedded planet and the aspect ratio of the disk produce changes in the level of contrast between $L_4$ and $L_5$. Systems with colder disks and those with more massive planets produce larger contrasts in the gas. 

%This is explained by the strong gravitational perturbation experienced by the co-orbital gas in these systems, as they are locally dominated by the planet.

%For systems with $m_{\rm p}\gtrsim150$ $M_{\oplus}$ and $h_{\rm p}\gtrsim0.09$ we observed a \cc{long-lasting plateau in the gas surface density evolution.} \cc{potential connection to observations}.

If the planet's mass is above $185$ $M_{\oplus}$, the gas evolution around the Lagrange points becomes chaotic, with overdensities in the outer gap perturbing the co-orbital gas, causing oscillations in its time evolution.

We also found deviations in the angular location of the peak near $L_4$ and $L_5$ with respect to the restricted three-body problem. This variation was well-characterized by the ratio between the planetary mass and the thermal mass, recovering the RTBP results in gravity-dominated systems ($Q\gg1)$. 

Finally, we discovered that the observed gas distribution around the Lagrange points is caused by the azimuthal variations in the radial velocity profile induced in the system by the disk-planet interaction.

\subsection{Beyond the non-migrating and circular planet}

Our results are based on a simple configuration where the perturbing planet is non-migrating and moving in a circular orbit around the central star. Therefore, in this section we relax those assumptions in order to investigate variations that may arise in the observed trend between $L_4$ and $L_5$.

\paragraph{Eccentricity} Taking our fiducial model with a negative temperature gradient $\beta=-0.1$, we ran 3 simulations with 3 different values for the planet's eccentricity: $0.5h_{\rm p}$, $h_{\rm p}$ and $1.5h_{\rm p}$. We found that in the first case, the time evolution of the gas distribution around $L_4$ and $L_5$ closely follows the blue curve in Figure \ref{fig:dens_evo}. However, when the eccentricity of the planet becomes significant, i.e. $e\gtrsim h_{\rm p}$, the relative asymmetry between $L_4$ and $L_5$ decreases by about 65\% in both amplitude and duration with respect to the blue curve in Figure \ref{fig:dens_evo}. Moreover, above eccentricities of $2h_{\rm p}$ the relative asymmetry is almost negligible. Figure 11 in \citealt{Rodenkirch2021} displays a similar trend, decreasing the amount of trapped material around the $L_5$ point when increasing the eccentricity of the planet.

\paragraph{Migration} For migration to have a significant effect on the co-orbital flow, the libration timescale, given by $\tau_{\rm lib}=8\pi r_{\rm p}/3\Omega_{\rm p}x_{\rm s}$, should be in the order of the migration timescale of the planet (i.e., $\tau_{\rm mig}\lesssim\tau_{\rm lib}$). For our fiducial model $\tau_{\rm lib}\sim15$ orbital periods, which is very small compared to the large migration timescales of gap-opening planets, thus, the libration motion of the gas is not perturbed by the planet moving inwards. Based on this, we do not expect planetary migration to change our results for typical parameters. To confirm this we performed one simulation with a planet migrating inwards over $\tau_{\rm mig}\sim10^4$ orbits and observed that the asymmetry prevails and does not change with respect to the expected blue curve in Figure \ref{fig:dens_evo}.

\paragraph{Mass Growth} To ensure that the trend we observe for the gas distribution around $L_4$ and $L_5$ is neither a transient effect nor an artifact caused by the instantaneous injection of planetary mass into the system. We ran a simulation with our fiducial parameters, a negative temperature gradient, and a smooth mass injection over 1000 orbits. We observed that when the mass taper is used, the gas distribution around the Lagrange points does not change and the time evolution resembles the blue curve in Figure \ref{fig:dens_evo}. The only effect the taper has in the system is that it shifts the curve in time, with evolution starting only after the planet has reached its final mass. The chaotic evolution driven by vortices triggered at the edge of the gap is influenced by the planet mass, here set to $m_{\rm p}=185$ $M_{\oplus}$. The timescale over which the planet grows can affect the lifetime of these vortices, as the planet grows more slowly, the vortices tend to become elongated and weaker \citep{Hammer2017,Hallam2020}, potentially shifting the onset of the chaotic regime to higher planet masses. However, \citet{Rometsch2021} showed that this effect is independent of the planet growth timescale, with the main difference being that vortices simply take longer to form.

\subsection{Comparison with previous work}

\cite{Montesinos2020} investigated the trapping of dust around Lagrange points. Although they found trends in the distribution of solids around $L_4$ and $L_5$, they worked with planetary masses considerably higher than those in our simulations, with Jupiter-mass planets and above (all of their systems are at the lower right end of Figure \ref{fig:sim_grid}). Based on our results, we would expect that the small dust particles ($\rm{St}\ll1$), which are well-coupled to the gas, may show some level of asymmetry. However, in their extreme regime, vortices driven by Rossby wave instability in the edges of the gap strongly disturb the co-orbital material, producing oscillations in the evolution of the particle distribution around $L_4$ and $L_5$. Furthermore, as these planets carve very deep gaps ($\log K >6.5$), the timescale on which an asymmetric gas distribution can form might be too short. Regarding the angular offset in the location of $L_4$ and $L_5$, their results are consistent with equation \ref{ec:offset}, noting that their super-thermal planets ($Q\gg1$) have particle distributions usually peaking at RTBP positions.

As a potential explanation for the crescent-shaped feature observed in the HD 163296 system, \cite{Rodenkirch2021} proposed a Jupiter-mass planet located at 48 au with respect to the central star. In their simulations, they found more accumulation of small dust particles around $L_5$ than $L_4$, which is consistent with our results for a negative temperature gradient in the disk (they use $\beta<0$). Additionally, the angular positions of $L_4$ and $L_5$ in their simulations agree with our results; for their fiducial parameters---$Q=4$---the azimuthal offset is expected to be almost zero.

Studying the formation of substructures in disks with compact planets, \cite{garrido2022} found that when vortices are formed in the co-orbital Lagrange points $L_4$ and $L_5$, they show a trend that is in agreement with our work. For a single-planet system and a negative temperature gradient, they observed that $L_5$ retains more gas than $L_4$. Moreover, the time evolution depicted in their Figure 6 is very similar to the one shown in Figure \ref{fig:dens_evo}, with a well-defined peak and a subsequent decay to zero.

\subsection{Application to observed systems}

There has been three cases of observed protoplanetary disks with tentative Lagrange point emission: PDS 70, HD 163296 and LkCa 15. 

\paragraph{PDS 70} This star is known to host two Jupiter-like planets in an internal cavity. \cite{Balsalobre-Ruza2023} showed that there is a highly significant emission at the putative $L_5$ point of planet b. The angular position of this emission agrees with the RTBP, which is consistent with the two super-thermal planets ($Q\gg1$) \citep{Wang+2020} having $|\phi_{\rm max}|\to60^{\circ}$ as predicted by Equation \ref{ec:offset}. Furthermore, if this detection is confirmed as a $L_5$ point we can claim, based on our results, that the temperature gradient in this disk must be negative.

\paragraph{HD 163296} \cite{Garrido-Deutelmoser2023} demonstrated that the crescent feature observed in this system could be produced by two gap-sharing planets, with the crescent being material accumulated at the $L_5$ location of the inner one. In that scenario, we would expect that the system has a negative temperature gradient. Moreover, the position of the center of this crescent, measured at $75^{\circ} \pm 10^{\circ}$ relative to the inner planet, aligns well with the prediction from equation \ref{ec:offset}, which gives a value of $\sim75^{\circ}$ for the $L_5$ point in the context of this planet.

\paragraph{LkCa 15} Although no planets have been confirmed in this system, \citealt{Long2022} identified robust non-axisymmetric dust continuum emission features within this disk. Specifically, they detected clump (potential $L_4$) and arc (potential $L_5$) morphologies that seem to be separated by roughly $120^{\circ}$ in azimuth along a faint ring located at 42 au with respect to the central star. Further follow-up observations of these features could potentially lead to the first-ever detection of an exoplanet based on emission from Lagrange points. Hydrodynamic simulations conducted by \cite{DongFung2017} suggested the presence of a potential planet with a mass ranging from 0.15 to 1.5 $M_{J}$, inferred from the gap morphology observed in scattered light. Combined with their derived value for $h_{\rm p}$ of 0.07, Equation \ref{ec:offset} gives an offset of $\lesssim18^{\circ}$ with respect to the RTBP for such a planet.

\paragraph{Future Observations} The Next-generation Very Large Array (ngVLA) will span a wavelength range of 2.5 to 300 mm and is expected to achieve an angular resolution of up to 0.1 mas at 2.5 mm (corresponding to 0.01 au at a distance of 100 pc). However, \citealt{Harter2020} tested the capabilities and showed a beam size of $\theta_{\rm res} = 0.2$ au at 3 mm wavelength. To resolve a dust-based Lagrange point, assuming a characteristic width of $w \approx 1.5 x_{\rm s}$, a size-to-resolution ratio of at least 1 is required ($w/\theta_{\rm res} > 1$). Under this criterion, a planet with mass $m_{\rm p} \geq 10\,M_{\oplus}$ located at $r_{\rm p} \geq 3\,\text{au}$, embedded in a disk with aspect ratio $h_{\rm p} \leq 0.03\,\left(r_{\rm p}/3 \,\rm{au}\right)^{1.8}$, fulfills the necessary conditions for detection around stars with luminosity $L_{\star} = 10\,L_{\odot}$, typical of Herbig Ae/Be stars. Favorable disk parameters include low viscosity $\alpha_{\rm ss} < 10^{-5}$ and surface density $\Sigma(r_{\rm p}) \leq 600\,\mathrm{g\,cm^{-2}}$, ensuring signal strength above ngVLA sensitivity limits \citep{Harter2020}. In contrast to ALMA, ngVLA is expected to detect larger dust grains, more similar to those observed with the Very Large Array (VLA). In the case of HD 163296, emission at 2.8 mm and 3.2 mm still reveals the Lagrangian point structure discussed previously. In these frequency bands, including those that overlap with the VLA, ngVLA will offer angular resolutions up to 15 times higher (see Figure 1 in \citealt{Guidi2022}).

\subsection{Tracing Dust}
The results found in this work only focus on the gas asymmetries; therefore, they can only be directly applied to small, well-coupled dust---with $\text{St}\ll1$. 

We ran a simulation with the fiducial parameters and a dust species in this low Stokes number regime to confirm our expectation, clearly observing that the accumulation at the Lagrange points follows the trend with the temperature gradient $\beta$. If anything, the dust displays structures that are significantly more pronounced than those observed in the gas. 

In the limit of largely decouples species ($\text{St}\gg1$), we expect that all the particles will mainly respond to the adiabatic changes from the orbital migration of the planets similar to \citet{Sicardy2003}. In turn, assessing the dynamical evolution of dusts species for intermediate values of $\text{St}$ require further analysis. We plan to further study the dynamical of dust species for a whide range of Stokes number in a subsequent work.

%As a second part of this work, we will investigate the asymmetries that might occur in the dust within the very decoupled regime, where $\text{St}\ll1$, and in the intermediate coupling regime, where $\text{St}\sim1$.

%\begin{acknowledgements}
%\emph{Acknowledgements} 
%\end{acknowledgements}
%\software{\textsc{Fargo3D}}

\section{Conclusions}

In this work, we find that the gas distribution around the co-orbital Lagrange points $L_4$ and $L_5$ of an embedded planet is solely regulated by the temperature gradient in the surrounding disk. In a globally isothermal disk, the gas is evenly distributed at these locations. However, when a temperature gradient is introduced, characterized by a positive or negative value of $\beta$, overdensities form at $L_4$ or $L_5$, respectively. 

As gas begins to accumulate near these Lagrange points, the azimuthal location of the peak in the gas deviates from the values predicted by the restricted three-body problem. We further characterize this trend using the ratio of the planet’s mass to its thermal mass, showing that the behavior approaches the RTBP result when the planet strongly dominates its vicinity ($Q\gg1$).  

When the disk has a temperature gradient, its interaction with the planet produces a subtle radial velocity profile with azimuthal variations across the protoplanetary disk. As a result, one of the libration regions around the Lagrange points expands more than the other, leading to an asymmetric distribution of co-orbital material.

Overall, we find that the overdensity around one of the Lagrange points is a robust outcome of disk-planet interaction. At lower-planetary masses ($m_{\rm p}<185$ $M_{\oplus}$) the longest-lived and largest-amplitude structures occur mainly for gap-opening planets with $\log K>2$ (i.e., with depths $\Sigma_{\rm gap}/\Sigma_0\lesssim 0.2$). At higher masses, the position and physical extent of the overdensity becomes time-variable, but generally sustained over 5000 orbits.

%paragraph with potential applications and future work (?)

\section*{Acknowledgments}

We thank Carolina Charalambous for her valuable support in setting up the N-body integrations and for her insightful discussions throughout the project.

We also thank Phil Armitage, Stanley Baronett, Konstantin Batygin, Fred Ciesla, Leonardo Krapp, Kaitlin Kratter, Feng Long, Diego Mu\~noz, Brandon Radzom, Antranik Sefilian and Songhu Wang for useful discussions and constructive feedback.

P.~B.~L. acknowledges  support from ANID, QUIMAL fund ASTRO21-0039 and FONDECYT project 1231205. C.~P. acknowledges support from ANID BASAL project FB210003.

\newpage
%\appendix
\renewcommand{\thefigure}{\thesection\arabic{figure}}
\setcounter{figure}{0}
\onecolumngrid

\bibliography{sample631}{}

\begin{thebibliography}{}
\expandafter\ifx\csname natexlab\endcsname\relax\def\natexlab#1{#1}\fi
\providecommand{\url}[1]{\href{#1}{#1}}
\providecommand{\dodoi}[1]{doi:~\href{http://doi.org/#1}{\nolinkurl{#1}}}
\providecommand{\doeprint}[1]{\href{http://ascl.net/#1}{\nolinkurl{http://ascl.net/#1}}}
\providecommand{\doarXiv}[1]{\href{https://arxiv.org/abs/#1}{\nolinkurl{https://arxiv.org/abs/#1}}}

\bibitem[{{Andrews}(2020)}]{Andrews2020}
{Andrews}, S.~M. 2020, \araa, 58, 483, \dodoi{10.1146/annurev-astro-031220-010302}

\bibitem[{{Bae} {et~al.}(2023){Bae}, {Isella}, {Zhu}, {Martin}, {Okuzumi}, \& {Suriano}}]{bae2023_PPPVII}
{Bae}, J., {Isella}, A., {Zhu}, Z., {et~al.} 2023, in Astronomical Society of the Pacific Conference Series, Vol. 534, Protostars and Planets VII, ed. S.~{Inutsuka}, Y.~{Aikawa}, T.~{Muto}, K.~{Tomida}, \& M.~{Tamura}, 423, \dodoi{10.48550/arXiv.2210.13314}

\bibitem[{{Balsalobre-Ruza, O.} {et~al.}(2023){Balsalobre-Ruza, O.}, {de Gregorio-Monsalvo, I.}, {Lillo-Box, J.}, {Hu\'elamo, N.}, {Ribas, \'A.}, {Benisty, M.}, {Bae, J.}, {Facchini, S.}, \& {Teague, R.}}]{Balsalobre-Ruza2023}
{Balsalobre-Ruza, O.}, {de Gregorio-Monsalvo, I.}, {Lillo-Box, J.}, {et~al.} 2023, A\&A, 675, A172, \dodoi{10.1051/0004-6361/202346493}

\bibitem[{{Benisty} {et~al.}(2021){Benisty}, {Bae}, {Facchini}, {Keppler}, {Teague}, {Isella}, {Kurtovic}, {P{\'e}rez}, {Sierra}, {Andrews}, {Carpenter}, {Czekala}, {Dominik}, {Henning}, {Menard}, {Pinilla}, \& {Zurlo}}]{Benisty2021}
{Benisty}, M., {Bae}, J., {Facchini}, S., {et~al.} 2021, \apjl, 916, L2, \dodoi{10.3847/2041-8213/ac0f83}

\bibitem[{Benítez-Llambay \& Masset(2016)}]{Benitez2016}
Benítez-Llambay, P., \& Masset, F.~S. 2016, The Astrophysical Journal Supplement Series, 223, 11.
\newblock \url{http://stacks.iop.org/0067-0049/223/i=1/a=11}

\bibitem[{{Chang} {et~al.}(2023){Chang}, {Youdin}, \& {Krapp}}]{chang23}
{Chang}, E., {Youdin}, A.~N., \& {Krapp}, L. 2023, \apjl, 946, L1

\bibitem[{{de Val-Borro} {et~al.}(2007){de Val-Borro}, {Artymowicz}, {D'Angelo}, \& {Peplinski}}]{de-val-borro07}
{de Val-Borro}, M., {Artymowicz}, P., {D'Angelo}, G., \& {Peplinski}, A. 2007, \aap, 471, 1043

\bibitem[{{Dong} \& {Fung}(2017)}]{DongFung2017}
{Dong}, R., \& {Fung}, J. 2017, \apj, 835, 146, \dodoi{10.3847/1538-4357/835/2/146}

\bibitem[{{Garrido-Deutelmoser} {et~al.}(2023){Garrido-Deutelmoser}, {Petrovich}, {Charalambous}, {Guzm{\'a}n}, \& {Zhang}}]{Garrido-Deutelmoser2023}
{Garrido-Deutelmoser}, J., {Petrovich}, C., {Charalambous}, C., {Guzm{\'a}n}, V.~V., \& {Zhang}, K. 2023, \apjl, 945, L37, \dodoi{10.3847/2041-8213/acbea8}

\bibitem[{{Garrido-Deutelmoser} {et~al.}(2022){Garrido-Deutelmoser}, {Petrovich}, {Krapp}, {Kratter}, \& {Dong}}]{garrido2022}
{Garrido-Deutelmoser}, J., {Petrovich}, C., {Krapp}, L., {Kratter}, K.~M., \& {Dong}, R. 2022, \apj, 932, 41, \dodoi{10.3847/1538-4357/ac6bfd}

\bibitem[{{Guidi} {et~al.}(2022){Guidi}, {Isella}, {Testi}, {Chandler}, {Liu}, {Schmid}, {Rosotti}, {Meng}, {Jennings}, {Williams}, {Carpenter}, {de Gregorio-Monsalvo}, {Li}, {Liu}, {Ortolani}, {Quanz}, {Ricci}, \& {Tazzari}}]{Guidi2022}
{Guidi}, G., {Isella}, A., {Testi}, L., {et~al.} 2022, \aap, 664, A137, \dodoi{10.1051/0004-6361/202142303}

\bibitem[{{Hallam} \& {Paardekooper}(2020)}]{Hallam2020}
{Hallam}, P.~D., \& {Paardekooper}, S.~J. 2020, \mnras, 491, 5759, \dodoi{10.1093/mnras/stz3437}

\bibitem[{{Hammer} {et~al.}(2017){Hammer}, {Kratter}, \& {Lin}}]{Hammer2017}
{Hammer}, M., {Kratter}, K.~M., \& {Lin}, M.-K. 2017, \mnras, 466, 3533, \dodoi{10.1093/mnras/stw3000}

\bibitem[{{Harter} {et~al.}(2020){Harter}, {Ricci}, {Zhang}, \& {Zhu}}]{Harter2020}
{Harter}, S.~K., {Ricci}, L., {Zhang}, S., \& {Zhu}, Z. 2020, \apj, 905, 24, \dodoi{10.3847/1538-4357/abcafc}

\bibitem[{{Isella} {et~al.}(2018){Isella}, {Huang}, {Andrews}, {Dullemond}, {Birnstiel}, {Zhang}, {Zhu}, {Guzm{\'a}n}, {P{\'e}rez}, {Bai}, {Benisty}, {Carpenter}, {Ricci}, \& {Wilner}}]{Isella2018}
{Isella}, A., {Huang}, J., {Andrews}, S.~M., {et~al.} 2018, \apjl, 869, L49, \dodoi{10.3847/2041-8213/aaf747}

\bibitem[{{Jim{\'e}nez} \& {Masset}(2017)}]{Jimenez2017}
{Jim{\'e}nez}, M.~A., \& {Masset}, F.~S. 2017, \mnras, 471, 4917, \dodoi{10.1093/mnras/stx1946}

\bibitem[{{Kanagawa} {et~al.}(2017){Kanagawa}, {Tanaka}, {Muto}, \& {Tanigawa}}]{Kanagawa2017}
{Kanagawa}, K.~D., {Tanaka}, H., {Muto}, T., \& {Tanigawa}, T. 2017, \pasj, 69, 97, \dodoi{10.1093/pasj/psx114}

\bibitem[{{Long} {et~al.}(2022{\natexlab{a}}){Long}, {Andrews}, {Zhang}, {Qi}, {Benisty}, {Facchini}, {Isella}, {Wilner}, {Bae}, {Huang}, {Loomis}, {{\"O}berg}, \& {Zhu}}]{Feng2022}
{Long}, F., {Andrews}, S.~M., {Zhang}, S., {et~al.} 2022{\natexlab{a}}, \apjl, 937, L1, \dodoi{10.3847/2041-8213/ac8b10}

\bibitem[{{Long} {et~al.}(2022{\natexlab{b}}){Long}, {Andrews}, {Zhang}, {Qi}, {Benisty}, {Facchini}, {Isella}, {Wilner}, {Bae}, {Huang}, {Loomis}, {{\"O}berg}, \& {Zhu}}]{Long2022}
---. 2022{\natexlab{b}}, \apjl, 937, L1, \dodoi{10.3847/2041-8213/ac8b10}

\bibitem[{{Lovelace} {et~al.}(1999){Lovelace}, {Li}, {Colgate}, \& {Nelson}}]{lovelace99}
{Lovelace}, R.~V.~E., {Li}, H., {Colgate}, S.~A., \& {Nelson}, A.~F. 1999, \apj, 513, 805

\bibitem[{{Lyra} {et~al.}(2009){Lyra}, {Johansen}, {Klahr}, \& {Piskunov}}]{Lyra2009}
{Lyra}, W., {Johansen}, A., {Klahr}, H., \& {Piskunov}, N. 2009, \aap, 493, 1125, \dodoi{10.1051/0004-6361:200810797}

\bibitem[{{Masset}(2000)}]{Masset2000}
{Masset}, F. 2000, Astronomy and Astrophysics Supplement Series, 141, 165, \dodoi{10.1051/aas:2000116}

\bibitem[{{Masset}(2002)}]{Masset2002}
{Masset}, F.~S. 2002, \aap, 387, 605, \dodoi{10.1051/0004-6361:20020240}

\bibitem[{{Montesinos} {et~al.}(2020){Montesinos}, {Garrido-Deutelmoser}, {Olofsson}, {Giuppone}, {Cuadra}, {Bayo}, {Sucerquia}, \& {Cuello}}]{Montesinos2020}
{Montesinos}, M., {Garrido-Deutelmoser}, J., {Olofsson}, J., {et~al.} 2020, \aap, 642, A224, \dodoi{10.1051/0004-6361/202038758}

\bibitem[{{M{\"u}ller} {et~al.}(2012){M{\"u}ller}, {Kley}, \& {Meru}}]{Muller2012}
{M{\"u}ller}, T.~W.~A., {Kley}, W., \& {Meru}, F. 2012, \aap, 541, A123, \dodoi{10.1051/0004-6361/201118737}

\bibitem[{{Murray}(1994)}]{Murray1994}
{Murray}, C.~D. 1994, \icarus, 112, 465, \dodoi{10.1006/icar.1994.1198}

\bibitem[{{Murray} \& {Dermott}(2000)}]{MD2000}
{Murray}, C.~D., \& {Dermott}, S.~F. 2000, {Solar System Dynamics} (Cambridge University Press)

\bibitem[{{Ogilvie} \& {Lubow}(2002)}]{Ogilvie+2002}
{Ogilvie}, G.~I., \& {Lubow}, S.~H. 2002, \mnras, 330, 950, \dodoi{10.1046/j.1365-8711.2002.05148.x}

\bibitem[{{Ogilvie} \& {Lubow}(2006)}]{Ogilvie2006}
---. 2006, \mnras, 370, 784, \dodoi{10.1111/j.1365-2966.2006.10506.x}

\bibitem[{{Perez} {et~al.}(2015){Perez}, {Dunhill}, {Casassus}, {Roman}, {Szul{\'a}gyi}, {Flores}, {Marino}, \& {Montesinos}}]{Perez2015}
{Perez}, S., {Dunhill}, A., {Casassus}, S., {et~al.} 2015, \apjl, 811, L5, \dodoi{10.1088/2041-8205/811/1/L5}

\bibitem[{{P{\'e}rez} {et~al.}(2020){P{\'e}rez}, {Casassus}, {Hales}, {Marino}, {Cheetham}, {Zurlo}, {Cieza}, {Dong}, {Alarc{\'o}n}, {Ben{\'\i}tez-Llambay}, {Fomalont}, \& {Avenhaus}}]{Perez2020}
{P{\'e}rez}, S., {Casassus}, S., {Hales}, A., {et~al.} 2020, \apjl, 889, L24, \dodoi{10.3847/2041-8213/ab6b2b}

\bibitem[{{Pinte} {et~al.}(2019){Pinte}, {van der Plas}, {M{\'e}nard}, {Price}, {Christiaens}, {Hill}, {Mentiplay}, {Ginski}, {Choquet}, {Boehler}, {Duch{\^e}ne}, {Perez}, \& {Casassus}}]{Pinte2019}
{Pinte}, C., {van der Plas}, G., {M{\'e}nard}, F., {et~al.} 2019, Nature Astronomy, 3, 1109, \dodoi{10.1038/s41550-019-0852-6}

\bibitem[{{Rodenkirch} {et~al.}(2021){Rodenkirch}, {Rometsch}, {Dullemond}, {Weber}, \& {Kley}}]{Rodenkirch2021}
{Rodenkirch}, P.~J., {Rometsch}, T., {Dullemond}, C.~P., {Weber}, P., \& {Kley}, W. 2021, \aap, 647, A174, \dodoi{10.1051/0004-6361/202038484}

\bibitem[{{Rometsch} {et~al.}(2021){Rometsch}, {Ziampras}, {Kley}, \& {B{\'e}thune}}]{Rometsch2021}
{Rometsch}, T., {Ziampras}, A., {Kley}, W., \& {B{\'e}thune}, W. 2021, \aap, 656, A130, \dodoi{10.1051/0004-6361/202142105}

\bibitem[{{Shakura} \& {Sunyaev}(1973)}]{Shakura1973}
{Shakura}, N.~I., \& {Sunyaev}, R.~A. 1973, \aap, 500, 33

\bibitem[{{Sicardy} \& {Dubois}(2003)}]{Sicardy2003}
{Sicardy}, B., \& {Dubois}, V. 2003, Celestial Mechanics and Dynamical Astronomy, 86, 321, \dodoi{10.1023/A:1024579912307}

\bibitem[{{Szul{\'a}gyi} {et~al.}(2018){Szul{\'a}gyi}, {Plas}, {Meyer}, {Pohl}, {Quanz}, {Mayer}, {Daemgen}, \& {Tamburello}}]{Szulagyi2018}
{Szul{\'a}gyi}, J., {Plas}, G. v.~d., {Meyer}, M.~R., {et~al.} 2018, \mnras, 473, 3573, \dodoi{10.1093/mnras/stx2602}

\bibitem[{{van der Marel} {et~al.}(2019){van der Marel}, {Dong}, {di Francesco}, {Williams}, \& {Tobin}}]{vanderMarel2019}
{van der Marel}, N., {Dong}, R., {di Francesco}, J., {Williams}, J.~P., \& {Tobin}, J. 2019, \apj, 872, 112, \dodoi{10.3847/1538-4357/aafd31}

\bibitem[{{Wang} {et~al.}(2020){Wang}, {Ginzburg}, {Ren}, {Wallack}, {Gao}, {Mawet}, {Bond}, {Cetre}, {Wizinowich}, {De Rosa}, {Ruane}, {Liu}, {Absil}, {Alvarez}, {Baranec}, {Choquet}, {Chun}, {Defr{\`e}re}, {Delorme}, {Duch{\^e}ne}, {Forsberg}, {Ghez}, {Guyon}, {Hall}, {Huby}, {Jolivet}, {Jensen-Clem}, {Jovanovic}, {Karlsson}, {Lilley}, {Matthews}, {M{\'e}nard}, {Meshkat}, {Millar-Blanchaer}, {Ngo}, {Orban de Xivry}, {Pinte}, {Ragland}, {Serabyn}, {Catal{\'a}n}, {Wang}, {Wetherell}, {Williams}, {Ygouf}, \& {Zuckerman}}]{Wang+2020}
{Wang}, J.~J., {Ginzburg}, S., {Ren}, B., {et~al.} 2020, \aj, 159, 263, \dodoi{10.3847/1538-3881/ab8aef}

\bibitem[{{Zhang} {et~al.}(2021){Zhang}, {Booth}, {Law}, {Bosman}, {Schwarz}, {Bergin}, {{\"O}berg}, {Andrews}, {Guzm{\'a}n}, {Walsh}, {Qi}, {van't Hoff}, {Long}, {Wilner}, {Huang}, {Czekala}, {Ilee}, {Cataldi}, {Bergner}, {Aikawa}, {Teague}, {Bae}, {Loomis}, {Calahan}, {Alarc{\'o}n}, {M{\'e}nard}, {Le Gal}, {Sierra}, {Yamato}, {Nomura}, {Tsukagoshi}, {P{\'e}rez}, {Trapman}, {Liu}, \& {Furuya}}]{zhang2021}
{Zhang}, K., {Booth}, A.~S., {Law}, C.~J., {et~al.} 2021, \apjs, 257, 5, \dodoi{10.3847/1538-4365/ac1580}

\bibitem[{{Zhang} {et~al.}(2018){Zhang}, {Zhu}, {Huang}, {Guzm{\'a}n}, {Andrews}, {Birnstiel}, {Dullemond}, {Carpenter}, {Isella}, {P{\'e}rez}, {Benisty}, {Wilner}, {Baruteau}, {Bai}, \& {Ricci}}]{Zhang2018}
{Zhang}, S., {Zhu}, Z., {Huang}, J., {et~al.} 2018, \apjl, 869, L47, \dodoi{10.3847/2041-8213/aaf744}

\end{thebibliography}
\bibliographystyle{aasjournal}
\end{document}